\RequirePackage{fix-cm}
\documentclass[smallextended]{svjour3}       
\smartqed  
\usepackage{graphicx}
\usepackage{mathptmx}      
\begin{document}

\title{Fundamental cosmology in the E-ELT era:}
\subtitle{The status and future role of tests of fundamental coupling stability}


\author{C.J.A.P. Martins}


\institute{C.J.A.P. Martins \at
              CAUP and IA-Porto, Rua das Estrelas s/n, 4150-762 Porto, Portugal \\
              Tel.: +351-226089891\\
              Fax: +351-226089831\\
              \email{Carlos.Martins@astro.up.pt}}          
\date{Received: date / Accepted: date}

\maketitle

\begin{abstract}
The observational evidence for the recent acceleration of the universe demonstrates that canonical theories of cosmology and particle physics are incomplete---if not incorrect---and that new physics is out there, waiting to be discovered. The most fundamental task for the next generation of astrophysical facilities is therefore to search for, identify and ultimately characterise this new physics. Here we highlight recent efforts along these lines, mostly focusing on ongoing work by CAUP's Dark Side Team aiming to develop some of the science case and optimise observational strategies for forthcoming facilities. The discussion is centred on tests of the stability of fundamental couplings (since the provide a direct handle on new physics), but synergies with other probes are also briefly considered. The goal is to show how a new generation of precision consistency tests of the standard paradigm will soon become possible.
\end{abstract}

\section{Introduction}
\label{intro}

In the middle of the XIX century Urbain Le Verrier and others mathematically discovered two new planets by insisting that the observed orbits of Uranus and Mercury agree with the predictions of Newtonian physics. The first of these---which we now call Neptune---was soon observed by Johann Galle and Heinrich d'Arrest. However, the second (dubbed Vulcan) was never found. We now know that the discrepancies in Mercury's orbit were a consequence of the fact that Newtonian physics can't adequately describe Mercury's orbit, and accounting for them was the first success of Einstein's General Relativity.

Over the past several decades, cosmologists have mathematically discovered two new components of the universe---which we have called dark matter and dark energy---but so far these have not been directly detected. Whether they will turn out to be Neptunes or Vulcans remains to be seen, but even their mathematical discovery alone highlights the fact that the standard $\Lambda$CDM paradigm, despite its phenomenological success, is at least incomplete.

Something similar applies to particle physics, where to some extent it is our confidence in the standard model that leads us to the expectation that there must be new physics beyond it. Neutrino masses, dark matter (again, assuming it exists) and the size of the baryon asymmetry of the universe all require new physics, and, significantly, all have obvious astrophysical and cosmological implications. Recent years have indeed made it clear that further progress in fundamental particle physics will increasingly depend on progress in cosmology.

Broadly speaking, fundamental physics can be defined to include two distinct but inter-related aspects \cite{Shaver}
\begin{itemize}
\item Tests of fundamental laws and symmetries, which includes tests of the Equivalence Principle (in its various forms), probing the behaviour of gravity on all scales, understanding the structure and dimensionality of spacetime, and testing the foundations of quantum mechanics; many of these principles are violated in extensions of the standard model.
\item Searches for fundamental constituents, including scalar fields as an explanation for dark energy, new particles for dark matter, magnetic monopoles or fundamental strings, as well as characterising the ones we already know (such as the Higgs, or the number and masses of neutrinos.
\end{itemize}

After a quest of several decades, the recent LHC evidence for a Higgs-like particle \cite{LHC1,LHC2} finally provides strong evidence in favour of the notion that fundamental scalar fields are part of Nature's building blocks. A pressing follow-up question is whether the associated field has some cosmological role, or indeed if there are additional scalar fields that do. As usual, the fact that we don't yet know the answer to the latter question has not prevented cosmologists and particle physicists from speculating, and indeed it may be a challenging to find one that has never used a scalar field at any point in his/her career.

Specifically, scalar fields play a key role in most paradigms of modern cosmology, including
\begin{itemize}
\item The period of exponential expansion of the early universe (inflation) that is believed to have seeded the density perturbations that led to the cosmic structures we now observe.
\item The dynamics of cosmological phase transitions and of their unavoidable relics (cosmic defects, such as strings, monopoles or domain walls).
\item Dynamical dark energy, and alternative to Einstein's cosmological constant for powering current acceleration phase (and, arguably, a more likely one).
\item The spacetime variation of nature's fundamental couplings, which is unavoidable in many extensions of the current standard model, and for which there is currently some tentative evidence.
\end{itemize}

Even more important than each of these paradigms is the fact that they don't occur alone: whenever a scalar field plays one of the above roles, it will also leave imprints in other contexts that one can look for. Although this complementary point is often overlooked, it will be crucial for the future of precision cosmology, since it can be exploited in the form of consistency tests of various paradigms. For example, in realistic models of inflation, the inflationary phase ends with a phase transition at which cosmic defects will form (and the energy scales of both will therefore be unavoidably related). More importantly, in realistic models of dark energy, where the dark energy is due to a dynamical scalar field, this field will naturally couple to the rest of the model (unless some unknown symmetry is postulated to suppress the couplings) and lead to potentially observable variations of nature's fundamental couplings.

In what follows we further develop the latter connection. We all know that fundamental couplings run with energy, and in many (or arguably most?) models they will equally naturally roll in time and ramble in space (meaning that they will depend on the local environment). Therefore astrophysical (and local) tests of their stability provide us with key probes of fundamental cosmology, and they can also (by themselves or in combination with other cosmological probes) shed light on the enigma of dark energy. The discussion mainly highlights ongoing work by CAUP's Dark Side Team aiming to develop some of the science case and optimise observational strategies for forthcoming facilities. In doing so we also summarise the contents of talks given by the author in the past few months at Moriond (joint cosmology/QCD session), the 2014 Azores cosmology school, MIAPP, ESO and NORDITA. Thus no attempts of completeness are made. The interested reader may find recent, more systematic reviews of related topics in \cite{Uzan,Weinberg,Amendola}.

\section{Fundamental couplings}
\label{consts}

Nature is characterised by a set of physical laws and fundamental dimensionless couplings, which historically we have assumed to be spacetime-invariant. For the former this is a cornerstone of the scientific method (it's hard to imagine how one could do science at all if it were not the case), but for the latter it is only a simplifying assumption without further justification. These couplings ultimately determine the properties of atoms, cells, planets and the universe as a whole, so it's remarkable how little we know about them. We have no 'theory of constants' that describes their role in physical theories or even which of them are really fundamental. If they do vary, all the physics we know is incomplete.

Fundamental couplings are indeed expected to vary in many extensions of the current standard model. In particular, this will be the case in theories with additional spacetime dimensions, such as string theory. In such paradigms the true fundamental constants are defined in higher dimensions, While the $(3+1)$-dimensional constants are effective quantities, typically related to the true constants via characteristic sizes of the extra dimensions. Many simple illustrations of these concepts exist, in Kaluza-Klein models \cite{Chodos}, superstring theories \cite{WuWang}, and brane world models \cite{Kiritsis}.

Interestingly, the first generation of string theorists had the hope that the theory would ultimately predict a unique set of laws and couplings for low-energy physics. However, following the discovery of the evidence for the acceleration of the universe this claim has been pragmatically replaced by an 'anything goes' approach, sometimes combined with anthropic arguments. Regardless of the merit of such approaches, is clear that experimental and observational tests of the stability of these couplings may well be their best route towards a testable prediction.

It goes without saying that a detection of varying fundamental couplings will be revolutionary: it will immediately prove that the Einstein Equivalence Principle is violated (and therefore that gravity can't be purely geometry) and that there is a fifth force of nature \cite{Damour}. But even improved null results are important and useful. The simple way to understand this is to realise that the natural scale for cosmological evolution of one of these couplings (driven by a fundamental scalar field) would be Hubble time. We would therefore expect a drift rate of the order of $10^{-10}$ per year. However, current local bounds, coming from atomic clock comparison experiments, are already orders of magnitude stronger \cite{Rosenband}.

Recent astrophysical evidence suggests, at more than four sigma level, a parts-per-million spatial variation of the fine-structure constant $\alpha$ at low redshifts \cite{Webb}. The data is unable to distinguish between a purely spatial dipole and one with an additional dependence on look-back time (both provide equally good fits to the data, at just above the four-sigma level). Although models that may explain such a result seem to require some amount of fine-tuning, it should also be said that there is also no identified systematic effect that can explain it, though some concerns have been recently raised \cite{Whitmore}. A possible class of models which may account for this type of spatial variations is that of symmetrons \cite{Marvin}, though a detailed analysis remains to be done. (Other possibilities in the literature rely on more canonical domain walls, but require significantly more fine-tuning, or at least properties of domain wall networks which certainly do not apply for standard domain walls.) One possible cause for caution (with these and other results) is that almost all of the existing data has been taken with other purposes in mind, whereas it is clear that this kind of measurements needs customised analysis pipelines and wavelength calibration procedures beyond those supplied by standard pipelines.

A recent joint analysis of all existing measurements indicates some inconsistencies \cite{FerreiraJiC}, and highlights the need for future more precise measurements. In the short term the PEPSI spectrograph at the LBT can play a role here, and in the longer term a new generation of high-resolution, ultra-stable spectrographs like ESPRESSO (for the VLT) and ELT-HIRES, which have these tests as a key science driver, will significantly improve the precision of these measurements (and crucially, have a much better control of systematics) and should be able to resolve the current controversy. (We will return to this point below.) A key technical improvement will be that ultimately one must do the wavelength calibration with laser frequency combs.

In theories where a dynamical scalar field yields a varying $\alpha$, the other gauge and Yukawa couplings are also expected to vary. In particular, in Grand Unified Theories the variation of $\alpha$ is related to that of energy scale of Quantum Chromodynamics, whence the nucleon masses necessarily vary when measured in an energy scale that is independent of QCD (such as the electron mass). It follows that we should expect a varying proton-to-electron mass ratio, $\mu=m_p/m_e$, which can be probed with $H_2$ \cite{Thompson} and other molecules.

Obviously, the specific relation between $\alpha(z)$ and $\mu(z)$ will be highly model-dependent, but this very fact makes this a unique discriminating tool between competing models. For example, in a broad class of unification scenarios \cite{Coc,Luo} one has
\begin{equation}
\frac{\Delta\mu}{\mu}=[0.8R-0.3(1+S)]\frac{\Delta\alpha}{\alpha}\,, \label{unification}
\end{equation}
where $R$ and $S$ are universal dimensionless parameters, respectively related to the strong and electroweak sectors of the model in question. Thus any model in this class may be phenomenologically characterised by its values of $R$ and $S$, and thus constrained using astrophysical measurements. It follows from this that it's highly desirable to identify astrophysical systems where various constants can be simultaneously measured, or systems where a constant can be measured in several independent ways. Systems where combinations of constants can be measured are also interesting, and can provide useful consistency tests \cite{Ferreira12,Ferreira13}.

Note that while for molecular hydrogen one is indeed measuring $\mu$, for more complex molecules (which may be more sensitive to $\mu$ variations than $H_2$ itself) one is actually measuring a ratio of an effective nucleon mass to the electron mass, and the relative variation of this quantity will only equal that of $\mu$ if there are no composition-dependent forces. A test of this hypothesis could thus by carried out by finding a system where $\mu$ can be separately measured from different molecules with different numbers of protons and neutrons---for example $H_2$, $HD$, ammonia and methanol.

\section{The UVES Large Programme}
\label{uveslp}

As mentioned in the previous section, one possible cause for caution regarding the results of \cite{Webb} was that it was based on archival data from the Keck and VLT telescopes, meaning that the data was originally taken for other purposes---by a large number of different observers, under a broad range of observing conditions, over a timespan of almost a decade---and subsequently re-analysed for this purpose. Thus although the dataset is quite large with 293 absorption systems in total, roughly half coming from each telescope (and a few observed by both), the data acquisition procedures were far form ideal, particularly regarding the key issue of wavelength calibration.

Trying to confirm these results was the main motivation for for the ongoing ESO UVES Large Programme (PI: Paolo Molaro). This is so far the only large program dedicated to tests of the stability of fundamental couplings, with an optimised sample and methodology. The programme consisted of about 40 VLT nights, in the period 2010-13, partly in service and partly in visitor mode. Key improvements in the data acquisition include obtaining calibration lamp exposures attached to science exposures (without resetting the cross-disperser encoding the position for each exposure) and observing bright (magnitude 9-11) asteroids at twilight, to monitor the radial velocity accuracy of UVES and the optical alignments. The collaboration includes members form all active observational groups.

With 40 VLT nights one can only observe a relatively small sample. Criteria for the sample selection included the presence of multiple absorption systems, brightness, relatively high redshift (so that the key FeII1608 transition is available), simplicity, narrow components at sensitive wavelengths, and no line broadening/saturation. Typically our resolution is around $R\sim60000$ and the signal-to-noise per pixel $S/N\sim100$. This led us to an expectation of a potential accuracy of $1-2$ parts-per-million (ppm) per system, where photon noise and calibration errors are comparable, and thus an overall goal of 2 ppm per system and 0.5 ppm for the full sample. Another of our key goals is to compare, check and optimise the different analysis pipelines available, including the introduction of blind analysis techniques.

The target selection was done before the dipole indications \cite{Webb} were known, and thus our sample is not optimised to test it (at least in the strict sense that we have no target near the north pole of the dipole). Our sample consists of 13 lines of sight for $\alpha$ measurements, and 2 lines of sight for $\mu$ measurements. Note that in the former case lines of sight often include several absorption systems, each of which may lead to a separate measurement. These are particularly useful for testing for hypothetical dependencies on look-back time. A brief description of our sample may be found in \cite{Bonifacio}. At the time of writing preliminary results on three of these lines of sight have already been published, an the analysis of the full sample is in progress. The raw data can already be found in the ESO public archive, and reduced data products will also be made public in due course.

The first complete quasar spectrum we analysed \cite{LP1} was that of HE 2217-2818, which includes 5 absorption systems at redshifts $z_{\rm abs} = 0.787$, 0.942, 1.556, 1.628 and 1.692. Applying the Many Multiplet method we found that the most precise result is obtained for the absorber at $z_{\rm abs} = 1.692$, where 3 Fe II transitions and Al II ${\lambda}$1670 have high S/N and provide a wide range of sensitivities to $\alpha$. Our final result for the relative variation in this system is
\begin{equation}
\frac{\Delta\alpha}{\alpha}=+1.3\pm2.4_{\rm stat}\pm{1.0}_{\rm sys}\, {\rm ppm}\,, \label{paperlp1}
\end{equation}
which is one of the tightest current bounds from an individual absorber. The absorbers towards quasar HE 2217-2818 reveal no evidence for variation in $\alpha$ at the 3 ppm precision level (with 1${\sigma}$ confidence). If the recently reported dipolar variation of $\alpha$ across the sky, with a maximum variation around 10ppm at the north pole of the dipole \cite{Webb}, were correct, the expectation at this sky position is $(3.2-5.4)\pm1.7$ ppm depending on dipole model used---specifically, depending on whether one assumes a pure spatial dipole or one with a further dependence on look-back time. Our constraint, indicated above, is not inconsistent with this expectation. 

We then carried out \cite{LP2} an accurate analysis of the $H_2$ absorption lines from the $z_{\rm abs}=2.402$ damped Ly${\alpha}$ system towards HE 0027-1836 to constrain the variation of$\mu$. A detailed cross-correlation analysis between 19 individual exposures, taken over three years, and the combined spectrum was carried out to check the wavelength calibration stability. We noticed the presence of possible wavelength dependent velocity drifts, and used available asteroid spectra taken with UVES close to our observations to confirm and quantify this effect. We use both linear regression analysis and Voigt profile fitting where ${\Delta}{\mu}/{\mu}$ is explicitly considered as an additional fitting parameter. Our final uncorrected result was ${\Delta}{\mu}/{\mu} = (-2.5 \pm 8.1_{\rm stat}\rm 6.2_{\rm sys})$ ppm, while when we applied the correction to the wavelength dependent velocity drift we found
\begin{equation}
\frac{\Delta\mu}{\mu}=-7.6\pm8.1_{\rm stat}\pm{6.3}_{\rm sys}\, {\rm ppm}\,; \label{paperlp2}
\end{equation}
the comparison of the two central values provides an indication of the potential importance of these effects, and highlights the need for further checks of previous results and for methods to account and correct for them in future observations. We note that intra-order and long-range distortions are not exclusive to the UVES spectrograph at the VLT, but have also been identified in HIRES at Keck and in HARPS. 

More recently \cite{LP3} we observed the equatorial quasar HS 1549$+$1919 with three telescopes: the Very Large Telescope, Keck and, for the first time in such analyses, Subaru. By directly comparing these spectra to each other, and by `supercalibrating' them using asteroid and iodine-cell tests, we detected and removed long-range distortions of the quasar spectra's wavelength scales which would have caused significant systematic errors in our $\alpha$ measurements. For each telescope we measure the relative deviation in $\alpha$ from the current laboratory value, $\Delta\alpha/\alpha$, in 3 absorption systems at redshifts $z_{\rm abs}=1.143$, 1.342, and 1.802. The nine measurements of $\Delta\alpha/\alpha$ are all consistent with zero at the 2-$\sigma$ level, with 1-$\sigma$ statistical (systematic) uncertainties in the range 5.6--24 (1.8--7.0) parts per million. They are also consistent with each other at the 1-$\sigma$ level, allowing us to form a combined value for each telescope and, finally, a single value for this line of sight:
\begin{equation}
\frac{\Delta\alpha}{\alpha}=-5.4\pm3.3_{\rm stat}\pm{1.5}_{\rm sys}\, {\rm ppm}\,, \label{paperlp3}
\end{equation}
which again is consistent with both zero and with the best-fit dipole predictions for this line of sight. If one averages all Large Programme $\alpha$ results so far, we obtain
\begin{equation}
\frac{\Delta\alpha}{\alpha}=-0.6\pm1.9_{\rm stat}\pm{0.9}_{\rm sys}\, {\rm ppm}\,. \label{paperlpall}
\end{equation}
Thus while a full analysis of our sample is still in progress, our results so far demonstrate the robustness and reliability at the 3 ppm level afforded by supercalibration techniques and the direct comparison of spectra from different telescopes. 

Our full sample consists of 15 among the brightest known quasars showing a suitable absorber, will more then 22 good absorption systems for which individual measurements can be made. On average we have therefore observed each absorber for more than three nights, which allowed us to build, for many of the absorbers, a much higher signal-to-noise ratio than achieved in all previous studies. In this case the photon statistical noise was reduced well below that from systematic errors. Our Large program achieved this for all relevant absorbers. We thus expect to have a signal-to-noise ratio that will allow us to detect, model and remove systematic errors down to the level of a few ppm. The final results of this analysis will be published in due course.

Before moving on to address the impact of these measurements on fundamental cosmology, let us pause to ask why these measurements are so difficult, and why the issue of systematics features quite frequently in the discussion. (Here we are focusing specifically on the spectroscopic measurements, as in the case of the previously discussed large Program.) The fact is that, while to some extent the radial velocity measurements in question are akin to finding exoplanets, they are  much harder in the present context, both because one is dealing with much fainter sources (QSOs with magnitude 16 or fainter, rather than bright stars) and because only a few absorption lines are clean enough.

In a nutshell, spectroscopic measurements of fundamental couplings require observing procedures - and instruments - beyond current facilities. Despite their success in other fields, spectrographs such as UVES, HARPS or Keck-HIRES were not built with this science case in mind and are far from optimal for it. One needs customised data reduction pipelines, as well as careful wavelength calibration procedures. In particular, one must calibrate with laser frequency combs \cite{Li,Steinmetz}, not rather than Th-Ar lamps or $I_2$. Fortunately, a new generation of high-resolution, ultra-stable spectrographs is forthcoming, which will have these measurements as key driver: shortly there will be PEPSI at LBT, in 2016 ESPRESSO at the VLT \cite{ESPRESSO}, and later on a high-resolution spectrograph at the E-ELT \cite{EELT1,EELT2}. At lower redshifts, there will also be complemented by ALMA measurements---two recent white papers discussing the ALMA role are \cite{ALMA1,ALMA2}.

\section{Aside: A bird's eye view of other measurements}
\label{otherobs}

While the Large Program results constitute the state-of-the-art in this field, they are complemented by a range of other local and astrophysical measurements, which probe the stability of fundamental couplings in a vast range of physical environments. While it is not the goal of this article to provide a thorough listing of all of these, in this section we provide some brief remarks on a few of these, on which there has been relevant recent activity. A more systematic review may be found in \cite{Uzan}.

In strict terms of sensitivity, the only other probe that is competitive with QSO spectroscopy is provided by laboratory tests using atomic clocks: in the case of $\alpha$, current sensitivity in the drift rate is \cite{Rosenband}
\begin{equation}
\frac{\dot\alpha}{\alpha}=(-1.6\pm2.3)\times10^{-17}\, {\rm yr}^{-1}\,; \label{localclock}
\end{equation}
for direct $\mu$ measurements the bounds are several orders of magnitude weaker. Nevertheless, different clock comparisons are sensitive to different combinations of $\alpha$, $\mu$ and the proton gyromagnetic ratio $g_p$, and a join analysis leads to useful constraints on the latter, as well as on some unification scenarios \cite{Ferreira12}. Significant progress is expected in laboratory measurements: with forthcoming molecular and nuclear clocks (particularly those based on Thorium229), a sensitivity as high as $10^{-21}$yr${^{-1}}$ may be achieved.

While direct measurements of $\alpha$ and $\mu$ are most commonly obtained in the ultraviolet/optical, in the radio band one can more often measure combinations of them. Typically combinations of HI 21cm absorption lines, conjugate 18cm OH lines and molecular rotation lines are sensitive to various combinations of $\alpha$, $\mu$ and $g_p$. Here a ppm sensitivity is nominally much easier to reach, though at significantly lower redshifts. A compilation of recent measurements, and a comparison with optical ones, may be found in \cite{FerreiraJiC}. The radio band sensitivity is even better within Galaxy ($z=0$), where one can search for environmental dependencies since measurements can be made in regions with densities that are many orders of magnitude smaller than the local one. Here the current best constraints come from \cite{Levshakov}, where no $\mu$ variation is seen at the 0.05 ppm level.

Compact objects have recently started to be explored as probes of the stability of fundamental couplings. This included both theoretical studies (studying the effect of varying couplings on these objects and using their known properties to infer {\it a posteriori} limits on such variations) and constraints obtained from direct observations. Here current sensitivities are around 50 ppm, and often the limiting factor comes from nuclear physics uncertainties. Theoretical analysis have been carried out for Population III stars \cite{Ekstrom}, solar-type stars \cite{Vieira} and neutron stars \cite{PerezGarcia}. More recently observational constraints have been obtained, at the above level of sensitivity, for both $\alpha$ and $\mu$ using neutron stars \cite{Berengut,Bagdonaite}: these come from spectroscopic observations of highly excited metal lines (FeV and NiV) and molecular hydrogen, respectively.

At higher redshift the cosmic microwave background provides a very clean probe in principle: varying couplings will obviously affect the ionisation history of the universe (including the energy levels, binding energies an Thomson cross-section). Nevertheless, the sensitivity of this probe is limited by the presence of degeneracies with other cosmological parameters, so current constraints are around the 2000 ppm level \cite{Planck}. Given the ppm constraints at low redshifts, CMB constraints will only be competitive for very specific classes of models that would predict strong variations in the very early universe---this would not be the case in the simplest dilaton-type (string-theory inspired) models. Moreover, current analysis assume that only $\alpha$ may vary (with all other couplings fixed) or at most also allow the electron mass to vary (with the caveats inherent to the discussion of a varying dimensionful coupling). A more realistic analysis, allowing both $\alpha$ and all particle masses to vary, is still pending. However, these studies do have a feature of interest, namely that they lead to constraints on the coupling between the putative scalar field and electromagnetism, independently (and on a completely different scale) from what is done in local tests, as illustrated in \cite{Erminia}.

At even higher redshifts constraints can be obtained from Big Bang nucleosynthesis, but they will necessarily be model-dependent. The first step in any analysis of the effect of varying fundamental couplings is to ascertain its effect on the neutron to proton mass difference, and this can only be done through the phenomenological Gasser-Leutwyler formula \cite{Gasser}. That said, current phenomenological constraints are at around the percent level for relatively generic models \cite{BBN}, though much tighter constraints can be obtained for more specific choices of model \cite{Coc}. Finally, it has been claimed that the Lithium problem might be removed in some GUT scenarios \cite{Stern}; a more detailed analysis is probably warranted given recent observational progress.

\section{Dynamical dark energy and varying couplings}
\label{darkside}

Observations suggest that the universe is dominated by an energy component whose gravitational behaviour is quite similar to that of a cosmological constant. Although a cosmological constant is consistent with existing data, its value would need to be so much smaller that particle physics expectations that a dynamical scalar field is arguably a more likely explanation. Such a field must be slow-rolling close to the present day (which is mandatory for $p<0$ and acceleration) and be dominating the dynamics, providing some $70\%$ or so of the critical density (which provides a rough normalisation). It then follows that \cite{Carroll} if the field couples to the rest of the model---{\bf which it will naturally do, unless some new symmetry is postulated to suppress the couplings}---it will lead to potentially observable long-range forces and time dependencies of the constants of nature.

Tests of the stability of fundamental couplings (whether they are detections of variations or null results) will constrain fundamental physics and cosmology. This therefore ensures a 'minimum guaranteed science': theoretical constraints will simply depend on the sensitivity of the tests. The importance of improved null results stems from the fact that there is no natural expectation for the scale of the putative variations, since they are controlled by an unknown parameter. But this also implies that any new, improved constraint will rule out some previously viable models. This is entirely analogous to cosmological constraints on dynamical dark energy: one is looking for deviations from the canonical $w_\phi=p_\phi/\rho_\phi=-1$, without any idea of when (meaning, at what level) or if such deviations will be found.

Scalar field based models for varying couplings can be divided into two broad classes. The first (which we simply refer to as Class I models) contains those where the degree of freedom responsible for the varying constants also provides the dark energy, These are therefore 'minimal' models, in the operational sense that there is a single new dynamical degree of freedom accounting for both. In this case the redshift dependence of the couplings is parametrically determined, and any available measurements (be they detections of null results) can be used to set constraints on combinations of the scalar field coupling and the dark energy equation of state. Specifically, the relative variation of $\alpha$ is given by \cite{Erminia}
\begin{equation}
\frac{\Delta\alpha}{\alpha}(z)=\zeta\int_0^z\sqrt{3\Omega_\phi(z')[1+w_\phi(z')]}\frac{dz'}{1+z'}\; \label{class1}
\end{equation}
here $\zeta$ is the dimensionless coupling of the scalar field to the electromagnetic sector of the theory, and $\Omega_\phi=\rho_\phi/(\rho_\phi+\rho_m)$ is the fraction of the energy density of the universe in dark energy. Examples of these models are discussed in \cite{Erminia,Rodger,SFS}. As is physically clear, if one sees no variations, either the field dynamics is very slow (ie, its equation of state is very close to $w=-1$) or the coupling is very small. In other words, astrophysical measurements mainly constrain the product of a cosmological parameter and a fundamental physics one. With a next-generation instrument such as ELT-HIRES one will either find variations or effectively rule out the simplest classes of these models, containing a single linearly coupled dynamical scalar field (unless one is prepared to accept very small---and thus fine-tuned---couplings $\zeta$).

However, this is not all. Standard observables such as supernovae are of limited use as dark energy probes \cite{Maor,Upadhye}, both because they probe relatively low redshifts (at least at the present time) and because to ultimately obtain the required cosmological parameters one effectively needs to take second derivatives of noisy data. A clear detection of varying $w(z)$ is crucial, given that we know that $w\sim -1$ today. Since the field is slow-rolling when dynamically important (close to the present day), a convincing detection of a varying w(z) will be tough at low redshift, and we must probe the deep matter era regime, where the dynamics of the hypothetical scalar field is fastest.

Varying fundamental couplings are ideal for probing scalar field dynamics beyond the domination regime \cite{Nunes}: such measurements can presently be made up to redshift $z=4.2$, and future facilities such as the E-ELT should be able to significantly extend this redshift range. Importantly, even null measurements of varying couplings can lead to interesting constraints on dark energy scenarios. Thus ALMA, ESPRESSO and ELT-HIRES can realise the prospect of a detailed characterisation of dark energy properties all the way until $z=4$ \cite{Amendola}, and possibly beyond. As we will see below, in the case of ELT-HIRES a reconstruction using quasar absorption lines can be more accurate than using supernova data, its key advantage being huge redshift lever arm. Importantly, these measurements have an additional key role: that of breaking degeneracies, when combined with more 'classical' probes, for constraining dynamical dark energy models. A case in point is that of ESA's Euclid mission, as was recently studied in \cite{Calabrese}. These degeneracies are broken not necessarily because measurements of varying couplings are intrinsically more constraining (that regime will only ensue for sufficiently large samples) but because the extended redshift lever arm effectively make is sensitive to different directions in the relevant parameter space.

Dark energy reconstruction using varying fundamental constants does assume that one is dealing with a Class I. As we will discuss later in this article there are various examples of modes for which this is not the case. Thus it is crucial for this analysis that in-built consistency checks exist, so that inconsistent assumptions can be identified and corrected. Explicit examples of incorrect assumptions that lead to observational inconsistencies can be found in \cite{Pauline1}. It precisely in closing the loop of consistency tests that the E-ELT will play the key role, particularly through the detection of the redshift drift signal \cite{Sandage} deep in the matter era, using the Ly-$\alpha$ forest and various additional metal absorption lines \cite{Liske}. The expected signal is
\begin{equation}
\frac{\Delta z}{\Delta t}=H_0(1+z)-H(z)\,, \label{zdrift}
\end{equation}
and this is a direct probe of the dynamics of the universe, without assumptions on gravity, geometry or clustering. It does not map out our (present-day) past light-cone, but directly measures evolution by comparing past light cones at different times. Therefore it provides an ideal probe of the dark sector in deep matter era. In practice the observable is the spectroscopic velocity
\begin{equation}
\frac{\Delta v}{v}=\frac{\Delta z}{1+z}\,. \label{specvel}
\end{equation}
The redshift drift is a key driver for ELT-HIRES, and possibly---at a fundamental level---ultimately the most important E-ELT deliverable. As shown in \cite{Pauline1,Pauline2}, having the ability to measure the stability of fundamental couplings and the redshift drift with a single instrument is a crucial strategic advantage.

Other facilities such as PEPSI at the LBT, the SKA \cite{Kloeckner}, ALMA  and intensity mapping experiments \cite{Yu} may also be able do measure the redshift drift. These will typically do it at lower redshifts. On the one hand these can therefore directly probe the accelerating phase of the universe (at redshifts that overlap with Euclid, for example), but on the other hand they will have a smaller lever arm---only the E-ELT can really probe the deep matter era, roughly $2<z<5$. Naturally the combination of low and high redshift measurements will lead to optimal constraints and will enable the discrimination between models that would otherwise be indistinguishable. In he case of the SKA, suggestions have been put forth to do it using neutral Hydrogen both at $z<1$ in emission and at $z>8$ in absorption. While the former should be easily within the reach of SKA-Phase 2, the latter will certainly be much harder.

Obviously, in addition to reconstructing the dark energy equation of state using fundamental couplings, supernovas and other cosmological observables will provide reconstructions at lower redshifts, so one can compare and combine the two reconstructions, as discussed in \cite{Amendola}. This can provide a check that the two reconstructions are consistent with each other (in the intermediate redshift range where the two datasets overlap), and assuming that they are consistent one can also infer a posterior likelihood for the coupling $\zeta$, since the fundamental couplings reconstructions depends on it but the one based on supernovas doesn't.

Finally, it should be pointed out that the another E-ELT instrument, ELT-IFU (in combination with JWST), should also dramatically extend the range of redshifts where cosmologically useful Type Ia supernovas are available---possibly up to a redshift $z\sim5$. A detailed assessment of the impact of these future datasets on fundamental cosmology is currently in progress. Interesting synergies are also expected to exist between these ground-based spectroscopic methods and Euclid, which need to be further explored.

\section{Case study: dark energy constraints}
\label{leite}

Here we provide a straw man analysis of how a reconstruction of the dark energy equation of state using measurements of the fine-structure constant $\alpha$ compares with a reconstruction using type Ia supernovas. The analysis mostly follows \cite{Leite}. We will base our theoretical analysis on PCA techniques, the formalism having been described in \cite{PCA}, to which we refer the reader for further details. One should bear in mind that PCA is a non-parametric method for constraining the dark energy equation of state. In assessing its performance, one should not compare it to parametric methods. Indeed, no such comparison is possible (even in principle), since the two methods are addressing different questions. Instead one should compare it with another non-parametric reconstruction, and for our purposes with varying couplings the type Ia supernovae provide a relevant comparison.

One can divide the relevant redshift range into $N_{\rm bin}$ bins such that in bin $i$ the equation of state parameter takes the value $w_i$, and the precision on the measurement of $w_i$ can be inferred from the Fisher matrix of the parameters $w_i$, specifically from $\sqrt{(F^{-1})_{ii}}$, and increases for larger redshift. One can however find a basis in which all the parameters are uncorrelated. This can be done by diagonalising the Fisher matrix such that $F = W^T \Lambda W$ where $\Lambda$ is diagonal and the rows of $W$ are the eigenvectors $e_i(z)$ or the principal components; the diagonal elements of are the eigenvalues $\lambda_i$ (ordered from largest to smallest) and define the variance of the new parameters, $\sigma^2_i = 1/\lambda_i$.

We will consider Class I quintessence type fields, with the simplest (linear) coupling to the electromagnetic sector. (This coupling will be marginalised over.) This can be seen as the first term of a Taylor expansion, and should be a good approximation if the field is slowly varying at low redshift. Then the evolution of $\alpha$ is as given above, and we will consider three fiducial forms for the equation of state parameter:
\begin{equation}
w_{c}(z)=-0.9,
\end{equation}
\begin{equation}
w_{s}(z)=-0.5+0.5 \tanh \left(z-1.5 \right),
\end{equation}
\begin{equation}
w_{b}(z)=-0.9+1.3 \exp{ \left[-\frac{(z-1.5)^2}{0.1} \right]}\,.
\end{equation}
At a phenomenological level, these describe the three qualitatively different interesting scenarios: an equation of state that remains close to a cosmological constant throughout the probed redshift range, one that evolves towards a matter-like behaviour by the highest redshifts probed, and one that has non-trivial features over a limited redshift range, perhaps associated to a low-redshift phase transition. Thus in what follows we will refer to these three cases as the \textit{constant}, \textit{step} and \textit{bump} fiducial models.

We will assume a flat universe, and further simplify the analysis by fixing $\Omega_m=0.3$. This is a standard procedure, and this specific choice of $\Omega_m$ has a negligible effect on the main result of the analysis, which is the uncertainty in the best determined modes---this has been discussed in \cite{PCA}. For each fiducial model we choose the coupling such that it leads to a few parts-per-million variation of $\alpha$ at redshift $z\sim4$, consistently with \cite{Webb}.

In order to systematically study possible observational strategies, one has to find an analytic expression for the behaviour of the uncertainties of the best determined PCA modes. For this one needs to explore the range of parameters such as the number of $\alpha$ measurements ($N_\alpha$) and the uncertainty in each measurement ($\sigma_\alpha$). For simplicity we will assume that this uncertainty is the same for each of the measurements in a given sample, and also that the measurements are uniformly distributed in the redshift range under consideration. This analysis is described in \cite{Leite}, which also discusses how to connect these theoretical tools to observational specifications. Specifically, using Monte Carlo techniques an overall normalisation (which can be described in terms of telescope time) can be derived from the present VLT performances, although one has to proceed with some caution since the present errors on $\alpha$ are dominated by systematics and not by photons.

Finally this calibrated formula can then be extrapolated to the expected performance of ESPRESSO and ELT-HIRES. In the former case, for a sample size optimised for the Guaranteed Time Observations one may foresee s factor of 3 gain (on average) in sensitivity due to improved signal-to-noise and resolution. These improvements arise from the fact that it will be, by design \cite{ESPRESSO}, free of the main systematics that are known to affect UVES, and in particular from the much more precise wavelength calibration, which will be done with a Laser Frequency Comb. Note that ESPRESSO does have a wavelength coverage that is substantially reduced compared to that of UVES, and this will certainly offset some of the improvements that would otherwise be achievable. In the latter case, \cite{EELT2} further gains in sensitivity are due to the five-fold increase in the telescope collecting area and to is its wide wavelength coverage, roughly matching UVES in the ultraviolet and optical but also going further into the infrared.

\begin{table}
\caption{\label{table1}Number of nights needed to achieve, with $\alpha$ measurements uniformly spaced in redshift, an uncertainty in the best-determined PCA mode equal to that expected from a SNAP-like dataset of 3000 Type Ia supernovas, for the ESPRESSO and ELT-HIRES spectrograph and the various fiducial models discussed in the text.}
\begin{center}
\begin{tabular}{|c|c|c|}
\hline
Model & ESPRESSO & ELT-HIRES \\
\hline
Constant & 649.8 & 19.5 \\
Step & 2231.6 & 66.9 \\
Bump & 1420.1 & 42.6 \\
\hline
\end{tabular}
\end{center}
\end{table}

We now assume 20 PCA bins and $\alpha$ measurements uniformly distributed in the redshift range $0.5<z<4.0$, and estimate the number of observation nights needed to obtain the same sensitivity on the first PCA mode as `classical' dataset of 3000 supernovas, assumed to be uniformly distributed up to $z\sim1.7$. Unsurprisingly we find that this is not possible at all with current UVES data (and the same should apply to current spectrographs at Keck or Subaru), while our estimates for ESPRESSO and ELT-HIRES are listed in Table \ref{table1}. We thus see that a few tens of nights are sufficient for ELT-HIRES, further highlighting the key role that the ELT will be able to play on fundamental cosmology.

For ESPRESSO, something like a thousand nights would be needed---not an impossible number as VLT time will become progressively 'cheaper' (and more focused on cutting-edge surveys) in the E-ELT era. In terms of cost, a back-of-the-envelope estimate would indicate comparable numbers in the two cases---something of order 60 MEuro, even including the cost of building a specific instrument. This is incomparably cheaper than any space-based facility. We note that a uniform redshift cover was important in obtaining the above results (and explains the different numbers for the three fiducial models). The range of redshifts considered also plays a role, as it will determine how many useful transitions will fall within the range of the spectrograph. A more detailed analysis allowing for these factors is in progress.

Our findings are directly relevant for the target selection process for both spectrographs, and even for the ELT-HIRES Phase A studies (which are due to star in about one year). The ELT-HIRES has clear potential for being a leading instrument in the field of fundamental cosmology. In addition to issues of resolution, stability and calibration, it is clear that a large redshift lever arm for the measurements is important, leading to the requirement of a broad wavelength range for the spectrograph (which also maximises the number of transitions available for the measurements).

\section{The quest for redundancy}
\label{redundancy}

Whichever way one eventually finds direct evidence for new physics, it will only be trusted once it is seen through multiple independent probes. This was manifest in the case of the discovery of the recent acceleration of the universe, where the supernova results were only accepted by the wider community once they were confirmed through CMB, large-scale structure and other data. It is clear that history will repeat itself in the case of varying fundamental couplings and/or dynamical dark energy. It is therefore crucial to develop consistency tests, in other words, astrophysical observables whose behaviour will also be non-standard as a consequence of either or both of the above.

An obvious example is that of violations of the Einstein Equivalence Principle. Varying fundamental couplings trivially violate Local Position Invariance, but one can also show \cite{Damour} that variations of $\alpha$ at few ppm level naturally lead to Weak Equivalence Principle violations within one order of magnitude of current bound on the Eotvos parameter. In that case an experiment such as MICRSOCOPE \cite{Microscope} should find these violations.

An astrophysical consistency test is provided by the comparison of the temperature-redshift relation and the distance duality (or Etherington) relation. The temperature-redshift relation is a robust prediction of standard cosmology, based on the assumptions of adiabatic expansion and photon number conservation, but it is violated in many scenarios, including string theory inspired ones. At a phenomenological level one can parametrise deviations to this law by adding an extra parameter, say $\beta$
\begin{equation}
T_{\rm CMB}=T_0(1+z)^{1-\beta}\,. \label{tofz}
\end{equation}
with current constraints on $\beta$ being around the $1.5\%$ level. (Note that here we're referring to direct constraints---indirect ones may also be inferred from spectral distortions \cite{Chluba}) Our recent work \cite{Tasos1} has shown that forthcoming data from Planck, ESPRESSO and ELT-HIRES will lead to much stronger constraints: Planck on its own can be as constraining as the existing bounds, ESPRESSO can improve on the current constraint by a factor of about three, and ELT-HIRES will improve on the current bound by one order or magnitude. We emphasise that estimates of all these gains rely on quite conservative on the number of sources (SZ clusters and absorption systems, respectively) where these measurements can be made. If the number of such sources increases, future constraints can be correspondingly stronger. Further improvements will come from proposed missions like COrE+ (a somewhat descoped version of \cite{Prism})

The distance duality relation is an equally robust prediction of standard cosmology; it assumes a metric theory of gravity and photon number conservation, but is violated if there's photon dimming, absorption or conversion. At a similarly phenomenological level one can parametrise deviations to this law by adding an extra parameter, say $\epsilon$
\begin{equation}
d_L=d_A(1+z)^{2+\epsilon}\,. \label{etheringt}
\end{equation}
with current constraints also being at the few percent level \cite{Tasos0}, and improvements are similarly expected from Euclid, the E-ELT and JWST.

In fact, in many models where photon number is not conserved---such as those where $\alpha$ varies---the temperature-redshift relation and the distance duality relation are not independent. Assuming adiabaticity and achromaticity one can in fact show that $\beta=-2\epsilon/3$, but it is easy to see that a direct relation exists for any such model. (A recent analysis in \cite{Hees}, with somewhat different assumptions, confirms our analysis.) This link allows one to use distance duality measurements to improve constraints on $\beta$, as first shown in \cite{Tasos1}.

Now, this relation provides an important consistency test for Class II models. These are the ones where the field that provides the varying couplings does not provide the dark energy (or at least does not provide all of it). In this case the link with dark energy is lost; if this was assumed in the context of an analysis as previously discussed (ie, assuming we were in the presence of a Class I model) it would lead to inconsistencies \cite{Pauline1}, but the temperature-distance duality relation could be used as a subsequent consistency test. Example of Class II models include Bekenstein-type models \cite{BSBM,Tasos2,Leal}.

This relation between $\alpha$ variations and the CMB temperature may be relevant, for example for Planck data analysis. If the ppm $\alpha$ dipole of \cite{Webb} is correct, then there should be a micro-Kelvin level dipole in the CMB temperature, in addition to the standard dipole due to our motion relative to the CMB frame. Note that even if in Class II model the degree of freedom that yields the varying couplings does not dominate the universe's dynamics at low redshift, it can bias cosmological parameter estimations, so constraining these variations is important for cosmological facilities. For example, in varying-$\alpha$ models the peak luminosity of Type Ia supernovas will depend on redshift, and samples of thousands of supernovas would be sensitive to ppm $\alpha$ variations.

Now, if photon number non-conservation changes observables such as $T(z)$, the distance duality relation, this may lead to additional biases, for example for Euclid. In \cite{Calabrese} we have quantified how these models weaken Euclid constraints on cosmological parameters, specifically those characterising the dark energy equation of state. Our results show that Euclid can, even on its own, constrain dark energy while allowing for photon number non-conservation. Naturally, stronger constraints can be obtained in combination with other probes. Interestingly, the ideal way to break a degeneracy involving the scalar-photon coupling is to use $T(z)$ measurements, which can be obtained with ALMA, ESPRESSO and ELT-HIRES (which, incidentally, may nicely complement each other in terms of redshift coverage for these measurements).

\section{Outlook}
\label{concl}

The observational evidence for the recent acceleration of the universe demonstrates that canonical theories of cosmology and particle physics are incomplete---if not incorrect---and that new physics is out there, waiting to be discovered. We have highlighted the key role that will be played by forthcoming high-resolution ultra-stable spectrographs in fundamental cosmology, by enabling a new generation of precision consistency tests. The most revolutionary among these is clearly the redshift drift, which is a key driver for ELT-HIRES, but may also be within the reach of other facilities, like PEPSI (at the LBT), SKA or even ALMA (although no sufficiently detailed studies exist for these at present).

Tests of the stability of fundamental couplings are crossing a threshold, with the first Large Program dedicated to them currently ongoing. So far everyone agrees that nothing is varying at the $10^{-5}$ (10 ppm) level out to redshifts $z\sim4$, with weaker constraints at higher redshifts and somewhat stronger ones within the galaxy $z\sim0$. Local tests with atomic clocks also provide very tight constraints. Note that a 10ppm constraint is already a very tight one (it's stronger than the Cassini bound on the Eddington PPN parameter, for example \cite{Cassini}), but improvements of 2-3 orders of magnitude may be foreseen in the coming years.

Together with the opportunity, afforded by astrophysical tests of the stability of fundamental couplings such as the fine-structure constant and the proton-to-electron mass ratio, to map and constrain additional dynamical degrees of freedom not only through the acceleration phase of the universe but also deep in the matter era (out to redshift $z\sim4$, and possibly beyond), these will consolidate this as an exciting new area of research, powered by dedicated new instruments.

Finally, let us again stress the role of consistency tests: taken together, tests of the stability of fundamental couplings, the redshift drift and constraints on the temperature-redshift relation and the distance duality relation will provide an exquisite mapping of the dark side of the universe. The E-ELT will enable further relevant tests, including tests of strong gravity around the galactic black hole, which were not discussed in this contribution. Last but not least, interesting synergies with other facilities, particularly ALMA, Euclid and the SKA, remain to be fully explored.

\begin{acknowledgements}
This work was done in the context of project PTDC/FIS/111725/2009, The Dark Side of the Universe, from FCT (Portugal). The author is also supported by an FCT Research Professorship, contract reference IF/00064/2012, funded by FCT/MCTES (Portugal) and POPH/FSE (EC).

Many interesting discussions with other members of CAUP's Dark Side team (Ana Catarina Leite, Ana Mafalda Monteiro, Jos\'e Pedro Vieira, Lu\'{\i}s Ventura, Mariana Juli\~ao, Marvin Silva, Miguel Ferreira, Pauline Vieizeuf, Pedro Leal and  Pedro Pedrosa) as well as with other collaborators in the work discussed herein, have shaped my views on this subject and are gratefully acknowledged.
\end{acknowledgements}

\bibliographystyle{spphys}       
\bibliography{martins}   

\begin{thebibliography}{10}
\providecommand{\url}[1]{{#1}}
\providecommand{\urlprefix}{URL }
\expandafter\ifx\csname urlstyle\endcsname\relax
  \providecommand{\doi}[1]{DOI \discretionary{}{}{}#1}\else
  \providecommand{\doi}{DOI \discretionary{}{}{}\begingroup
  \urlstyle{rm}\Url}\fi

\bibitem{Shaver}
P.~Shaver (ed.).
\newblock \emph{{Astronomy, cosmology and fundamental physics. Proceedings,
  ESO/CERN/ESA Symposium, Garching, Germany, March 4-7, 2002}} (2003)

\bibitem{LHC1}
G.~Aad, et~al., Phys.Lett. \textbf{B716}, 1 (2012).
\newblock \doi{10.1016/j.physletb.2012.08.020}

\bibitem{LHC2}
S.~Chatrchyan, et~al., Phys.Lett. \textbf{B716}, 30 (2012).
\newblock \doi{10.1016/j.physletb.2012.08.021}

\bibitem{Uzan}
J.P. Uzan, Living Rev.Rel. \textbf{14}, 2 (2011)

\bibitem{Weinberg}
D.H. Weinberg, M.J. Mortonson, D.J. Eisenstein, C.~Hirata, A.G. Riess, et~al.,
  Phys.Rept. \textbf{530}, 87 (2013).
\newblock \doi{10.1016/j.physrep.2013.05.001}

\bibitem{Amendola}
L.~Amendola, et~al., Living Rev.Rel. \textbf{16}, 6 (2013)

\bibitem{Chodos}
A.~Chodos, S.L. Detweiler, Phys.Rev. \textbf{D21}, 2167 (1980).
\newblock \doi{10.1103/PhysRevD.21.2167}

\bibitem{WuWang}
Y.S. Wu, Z.~Wang, Phys.Rev.Lett. \textbf{57}, 1978 (1986).
\newblock \doi{10.1103/PhysRevLett.57.1978}

\bibitem{Kiritsis}
E.~Kiritsis, JHEP \textbf{9910}, 010 (1999).
\newblock \doi{10.1088/1126-6708/1999/10/010}

\bibitem{Damour}
T.~Damour, Astrophys.Space Sci. \textbf{283}, 445 (2003).
\newblock \doi{10.1023/A:1022596316014}

\bibitem{Rosenband}
T.~Rosenband, D.~Hume, P.~Schmidt, C.~Chou, A.~Brusch, L.~Lorini, W.~Oskay,
  R.~Drullinger, T.~Fortier, J.~Stalnaker, S.~Diddams, W.~Swann, N.~Newbury,
  W.~Itano, D.~Wineland, J.~Bergquist, Science \textbf{319}, 1808 (2008).
\newblock \doi{10.1126/science.1154622}

\bibitem{Webb}
J.~Webb, J.~King, M.~Murphy, V.~Flambaum, R.~Carswell, et~al., Phys.Rev.Lett.
  \textbf{107}, 191101 (2011).
\newblock \doi{10.1103/PhysRevLett.107.191101}

\bibitem{Whitmore}
J.B. Whitmore, M.T. Murphy, {Impact of instrumental systematic errors on
  fine-structure constant measurements with quasar spectra} (2014).
\newblock Eprint arXiv:1409.4467

\bibitem{Marvin}
M.F. Silva, H.A. Winther, D.F. Mota, C.J.A.P. Martins, Phys.Rev. \textbf{D89},
  024025 (2014).
\newblock \doi{10.1103/PhysRevD.89.024025}

\bibitem{FerreiraJiC}
M.C. Ferreira, O.~Frigola, C.J.A.P. Martins, A.M.R.V.L. Monteiro, J.~Sol\`a,
  Phys.Rev. \textbf{D89}, 083011 (2014).
\newblock \doi{10.1103/PhysRevD.89.083011}

\bibitem{Thompson}
R.I. {Thompson}, Astrophys.Lett. \textbf{16}, 3 (1975)

\bibitem{Coc}
A.~Coc, N.J. Nunes, K.A. Olive, J.P. Uzan, E.~Vangioni, Phys.Rev. \textbf{D76},
  023511 (2007).
\newblock \doi{10.1103/PhysRevD.76.023511}

\bibitem{Luo}
F.~Luo, K.A. Olive, J.P. Uzan, Phys.Rev. \textbf{D84}, 096004 (2011).
\newblock \doi{10.1103/PhysRevD.84.096004}

\bibitem{Ferreira12}
M.C. Ferreira, M.D. Juli\~ao, C.J.A.P. Martins, A.M.R.V.L. Monteiro, Phys.Rev.
  \textbf{D86}, 125025 (2012).
\newblock \doi{10.1103/PhysRevD.86.125025}

\bibitem{Ferreira13}
M.C. Ferreira, M.D. Juli\~ao, C.J.A.P. Martins, A.M.R.V.L. Monteiro, Phys.Lett.
  \textbf{B724}, 1 (2013).
\newblock \doi{10.1016/j.physletb.2013.05.055}

\bibitem{Bonifacio}
P.~{Bonifacio}, H.~{Rahmani}, J.B. {Whitmore}, M.~{Wendt}, M.~{Centurion},
  P.~{Molaro}, R.~{Srianand}, M.T. {Murphy}, P.~{Petitjean}, I.I. {Agafonova},
  S.~{D'Odorico}, T.M. {Evans}, S.A. {Levshakov}, S.~{Lopez}, C.J.A.P.
  {Martins}, D.~{Reimers}, G.~{Vladilo}, Astronomische Nachrichten
  \textbf{335}, 83 (2014).
\newblock \doi{10.1002/asna.201312005}

\bibitem{LP1}
P.~Molaro, M.~Centurion, J.~Whitmore, T.~Evans, M.~Murphy, et~al.,
  Astron.Astrophys. \textbf{555}, A68 (2013).
\newblock \doi{10.1051/0004-6361/201321351}

\bibitem{LP2}
H.~Rahmani, M.~Wendt, R.~Srianand, P.~Noterdaeme, P.~Petitjean, et~al.,
  Mon.Not.Roy.Astron.Soc. \textbf{435}, 861 (2013).
\newblock \doi{10.1093/mnras/stt1356}

\bibitem{LP3}
T.M. {Evans}, M.T. {Murphy}, J.B. {Whitmore}, T.~{Misawa}, M.~{Centurion},
  S.~{D'Odorico}, S.~{Lopez}, C.J.A.P. {Martins}, P.~{Molaro}, P.~{Petitjean},
  H.~{Rahmani}, R.~{Srianand}, M.~{Wendt}, Mon.Not.Roy.Astron.Soc.
  \textbf{445}, 128 (2014).
\newblock \doi{10.1093/mnras/stu1754}

\bibitem{Li}
C.H. Li, A.J. Benedick, P.~Fendel, A.G. Glenday, F.X. Kaertner, et~al., Nature
  \textbf{452}, 610 (2008).
\newblock \doi{10.1038/nature06854}

\bibitem{Steinmetz}
T.~Steinmetz, T.~Wilken, C.~Araujo-Hauck, R.~Holzwarth, T.W. Hansch, et~al.,
  Science \textbf{321}, 1335 (2008).
\newblock \doi{10.1126/science.1161030}

\bibitem{ESPRESSO}
F.~{Pepe}, S.~{Cristiani}, R.~{Rebolo}, N.C. {Santos}, H.~{Dekker},
  D.~{M{\'e}gevand}, F.M. {Zerbi}, A.~{Cabral}, P.~{Molaro}, P.~{Di
  Marcantonio}, M.~{Abreu}, M.~{Affolter}, M.~{Aliverti}, C.~{Allende Prieto},
  M.~{Amate}, G.~{Avila}, V.~{Baldini}, P.~{Bristow}, C.~{Broeg}, R.~{Cirami},
  J.~{Coelho}, P.~{Conconi}, I.~{Coretti}, G.~{Cupani}, V.~{D'Odorico}, V.~{De
  Caprio}, B.~{Delabre}, R.~{Dorn}, P.~{Figueira}, A.~{Fragoso}, S.~{Galeotta},
  L.~{Genolet}, R.~{Gomes}, J.I. {Gonz{\'a}lez Hern{\'a}ndez}, I.~{Hughes},
  O.~{Iwert}, F.~{Kerber}, M.~{Landoni}, J.L. {Lizon}, C.~{Lovis}, C.~{Maire},
  M.~{Mannetta}, C.~{Martins}, M.A. {Monteiro}, A.~{Oliveira}, E.~{Poretti},
  J.L. {Rasilla}, M.~{Riva}, S.~{Santana Tschudi}, P.~{Santos}, D.~{Sosnowska},
  S.~{Sousa}, P.~{Span{\`o}}, F.~{Tenegi}, G.~{Toso}, E.~{Vanzella}, M.~{Viel},
  M.R. {Zapatero Osorio}, The Messenger \textbf{153}, 6 (2013)

\bibitem{EELT1}
ESO, {The E-ELT Construction Proposal} (2011)

\bibitem{EELT2}
R.~Maiolino, M.~Haehnelt, M.~Murphy, D.~Queloz, L.~Origlia, et~al., {A
  Community Science Case for E-ELT HIRES} (2013)

\bibitem{ALMA1}
V.~Fish, W.~Alef, J.~Anderson, K.~Asada, A.~Baudry, et~al.,
  {High-Angular-Resolution and High-Sensitivity Science Enabled by Beamformed
  ALMA} (2013)

\bibitem{ALMA2}
R.~Tilanus, T.~Krichbaum, J.~Zensus, A.~Baudry, M.~Bremer, et~al., {Future
  mmVLBI Research with ALMA: A European vision} (2014)

\bibitem{Levshakov}
S.~Levshakov, D.~Reimers, C.~Henkel, B.~Winkel, A.~Mignano, et~al.,
  Astron.Astrophys. \textbf{559}, A91 (2013).
\newblock \doi{10.1051/0004-6361/201322535}

\bibitem{Ekstrom}
S.~{Ekstr{\"o}m}, A.~{Coc}, P.~{Descouvemont}, G.~{Meynet}, K.A. {Olive}, J.P.
  {Uzan}, E.~{Vangioni}, A. \& A. \textbf{514}, A62 (2010).
\newblock \doi{10.1051/0004-6361/200913684}

\bibitem{Vieira}
J.P.P. Vieira, C.J.A.P. Martins, M.J.P.F.G. Monteiro, Phys.Rev. \textbf{D86},
  043003 (2012).
\newblock \doi{10.1103/PhysRevD.86.043003}

\bibitem{PerezGarcia}
M.A. Perez-Garcia, C.J.A.P. Martins, Phys.Lett. \textbf{B718}, 241 (2012).
\newblock \doi{10.1016/j.physletb.2012.10.047}

\bibitem{Berengut}
J.~Berengut, V.~Flambaum, A.~Ong, J.~Webb, J.D. Barrow, et~al., Phys.Rev.Lett.
  \textbf{111}(1), 010801 (2013).
\newblock \doi{10.1103/PhysRevLett.111.010801}

\bibitem{Bagdonaite}
J.~Bagdonaite, E.J. Salumbides, S.P. Preval, M.A. Barstow, J.D. Barrow, et~al.,
  {Limits on a Gravitational Field Dependence of the Proton--Electron Mass
  Ratio from H$_2$ in White Dwarf Stars} (2014).
\newblock \doi{10.1103/PhysRevLett.113.123002}

\bibitem{Planck}
P.~Ade, et~al., {Planck intermediate results. XXIV. Constraints on variation of
  fundamental constants} (2014)

\bibitem{Erminia}
E.~Calabrese, E.~Menegoni, C.J.A.P. Martins, A.~Melchiorri, G.~Rocha, Phys.Rev.
  \textbf{D84}, 023518 (2011).
\newblock \doi{10.1103/PhysRevD.84.023518}

\bibitem{Gasser}
J.~Gasser, H.~Leutwyler, Nucl.Phys. \textbf{B307}, 763 (1988).
\newblock \doi{10.1016/0550-3213(88)90107-1}

\bibitem{BBN}
C.J.A.P. Martins, E.~Menegoni, S.~Galli, G.~Mangano, A.~Melchiorri, Phys.Rev.
  \textbf{D82}, 023532 (2010).
\newblock \doi{10.1103/PhysRevD.82.023532}

\bibitem{Stern}
S.~Stern, {Dynamical dark energy and variation of fundamental constants (PhD
  Thesis)} (2008)

\bibitem{Carroll}
S.M. Carroll, Phys.Rev.Lett. \textbf{81}, 3067 (1998).
\newblock \doi{10.1103/PhysRevLett.81.3067}

\bibitem{Rodger}
R.I. Thompson, C.J.A.P. Martins, P.E. Vielzeuf, Mon.Not.Roy.Astron.Soc.
  \textbf{428}, 2232 (2013).
\newblock \doi{10.1093/mnras/sts187}

\bibitem{SFS}
M.P. {D{\c a}browski}, T.~{Denkiewicz}, C.J.A.P. {Martins}, P.E. {Vielzeuf},
  Phys.Rev. (12), 123512 (2014).
\newblock \doi{10.1103/PhysRevD.89.123512}

\bibitem{Maor}
I.~Maor, R.~Brustein, P.J. Steinhardt, Phys.Rev.Lett. \textbf{86}, 6 (2001).
\newblock \doi{10.1103/PhysRevLett.86.6}

\bibitem{Upadhye}
A.~Upadhye, M.~Ishak, P.J. Steinhardt, Phys.Rev. \textbf{D72}, 063501 (2005).
\newblock \doi{10.1103/PhysRevD.72.063501}

\bibitem{Nunes}
N.J. Nunes, J.E. Lidsey, Phys.Rev. \textbf{D69}, 123511 (2004).
\newblock \doi{10.1103/PhysRevD.69.123511}

\bibitem{Calabrese}
E.~{Calabrese}, M.~{Martinelli}, S.~{Pandolfi}, V.F. {Cardone}, C.J.A.P.
  {Martins}, S.~{Spiro}, P.E. {Vielzeuf}, Phys.Rev. (8), 083509 (2014).
\newblock \doi{10.1103/PhysRevD.89.083509}

\bibitem{Pauline1}
P.E. Vielzeuf, C.J.A.P. Martins, Phys.Rev. \textbf{D85}, 087301 (2012).
\newblock \doi{10.1103/PhysRevD.85.087301}

\bibitem{Sandage}
A.~{Sandage}, Astrophys. J. \textbf{136}, 319 (1962).
\newblock \doi{10.1086/147385}

\bibitem{Liske}
J.~Liske, A.~Grazian, E.~Vanzella, M.~Dessauges, M.~Viel, et~al.,
  Mon.Not.Roy.Astron.Soc. \textbf{386}, 1192 (2008).
\newblock \doi{10.1111/j.1365-2966.2008.13090.x}

\bibitem{Pauline2}
M.~Martinelli, S.~Pandolfi, C.J.A.P. Martins, P.E. Vielzeuf, Phys.Rev.
  \textbf{D86}, 123001 (2012).
\newblock \doi{10.1103/PhysRevD.86.123001}

\bibitem{Kloeckner}
F.~Aharonian, T.~Arshakian, B.~Allen, R.~Banerjee, R.~Beck, et~al., {Pathway to
  the Square Kilometre Array - The German White Paper -} (2013)

\bibitem{Yu}
H.R. Yu, T.J. Zhang, U.L. Pen, Phys.Rev.Lett. \textbf{113}, 041303 (2014).
\newblock \doi{10.1103/PhysRevLett.113.041303}

\bibitem{Leite}
A.C.O. Leite, C.J.A.P. Martins, P.O.J. Pedrosa, N.J. Nunes, Phys.Rev.
  \textbf{D90}, 063519 (2014)

\bibitem{PCA}
L.~Amendola, A.C.O. Leite, C.J.A.P. Martins, N.~Nunes, P.O.J. Pedrosa, et~al.,
  Phys.Rev. \textbf{D86}, 063515 (2012).
\newblock \doi{10.1103/PhysRevD.86.063515}

\bibitem{Microscope}
P.~Touboul, G.~Metris, V.~Lebat, A.~Robert, Class.Quant.Grav. \textbf{29},
  184010 (2012).
\newblock \doi{10.1088/0264-9381/29/18/184010}

\bibitem{Chluba}
J.~{Chluba}, M.N.R.A.S. \textbf{443}, 1881 (2014).
\newblock \doi{10.1093/mnras/stu1260}

\bibitem{Tasos1}
A.~Avgoustidis, G.~Luzzi, C.J.A.P. Martins, A.M.R.V.L. Monteiro, JCAP
  \textbf{1202}, 013 (2012).
\newblock \doi{10.1088/1475-7516/2012/02/013}

\bibitem{Prism}
P.~Andr\'e, et~al., JCAP \textbf{1402}, 006 (2014).
\newblock \doi{10.1088/1475-7516/2014/02/006}

\bibitem{Tasos0}
A.~Avgoustidis, C.~Burrage, J.~Redondo, L.~Verde, R.~Jimenez, JCAP
  \textbf{1010}, 024 (2010).
\newblock \doi{10.1088/1475-7516/2010/10/024}

\bibitem{Hees}
A.~Hees, O.~Minazzoli, J.~Larena, {On a breaking of the equivalence principle
  in the electromagnetic sector and its cosmological signatures} (2014)

\bibitem{BSBM}
H.B. Sandvik, J.D. Barrow, J.~Magueijo, Phys.Rev.Lett. \textbf{88}, 031302
  (2002).
\newblock \doi{10.1103/PhysRevLett.88.031302}

\bibitem{Tasos2}
A.~Avgoustidis, C.J.A.P. Martins, A.M.R.V.L. Monteiro, P.E. Vielzeuf, G.~Luzzi,
  JCAP \textbf{1406}, 062 (2014).
\newblock \doi{10.1088/1475-7516/2014/06/062}

\bibitem{Leal}
P.M.M. Leal, C.J.A.P. Martins, L.B. Ventura, Phys.Rev. \textbf{D90}, 027305
  (2014).
\newblock \doi{10.1103/PhysRevD.90.027305}

\bibitem{Cassini}
B.~Bertotti, L.~Iess, P.~Tortora, Nature \textbf{425}, 374 (2003).
\newblock \doi{10.1038/nature01997}

\end{thebibliography}

%
%

\end{document}